\documentstyle[12pt,epsfig]{article}
\begin{document}
\baselineskip 8mm
\title{\bf {\it Awaking} and {\it Sleeping} a Complex Network}

\author{ \bf R. L\'{o}pez-Ruiz$^{\ddag}$, Y. Moreno$^{\dag}$, A. F. Pacheco$^{\dag}$\\
 \small $^{\ddag}$ Departament of Computer Science and BIFI, \\
 \small $^{\dag}$ Departament of Theoretical Physics and BIFI, \\
 \small Faculty of Sciences - University of Zaragoza, \\
 \small 50009 - Zaragoza (Spain). \\
          \\
 \bf S. Boccaletti$^{\S}$ and D.-U. Hwang$^{\S}$  \\
 \small $^{\S}$ Istituto Nazionale di Ottica Applicata \\ 
 \small Largo Enrico Fermi, 6, 50125-Firenze (Italy) }
 
 \date{ }

\maketitle

\begin{center} {\bf Abstract} \end{center}
A network with local dynamics of logistic type is considered.  We
implement a mean-field multiplicative coupling among first-neighbor
nodes.  When the coupling parameter is small the dynamics is
dissipated and there is no activity: the network is {\it turned
off}. For a critical value of the coupling a non-null stable
synchronized state, which represents a {\it turned on} network,
emerges.  This global bifurcation is independent of the network
topology.  We characterize the bistability of the system by studying
how to perform the transition, which now is topology dependent, from
the active state to that with no activity, for the particular case of
a scale free network. This could be a naive model for the {\it
wakening} and {\it sleeping} of a brain-like system.

{\small
{\bf Electronic mails:} $^{\ddag}$ rilopez@unizar.es ; $^{\dag}$
yamir@unizar.es, amalio@unizar.es ; $^{\S}$ stefano@ino.it, duhwang@ino.it

\newpage

\section{Introduction}
%\label{sec:intro}

Understanding the brain is a formidable challenge. A multidisciplinary effort is
required to enlighten how it works, how it processes information and
how it takes decisions.  Furthermore, the brain behaviors which are considered
far from those commonly accepted, that is, the mental illnesses or the
degenerative cerebral evolutions, are important problems that are
under constant investigation.  The different approaches from the most
diverse fields, i.e., medicine, psychiatry, neurobiology, chemistry
and neural computation, should combine their particular visions
for trying to reach the collective goal of creating an artificial
brain-like system or, at least, in order to reach solutions to the most
diverse dysfunction symptoms which are found in its behavior.

Different models have been proposed to catch the computational
principles of mental processes. Neural networks are considered as a
paradigmatic model alternative to the more traditional models such as
finite automata, Turing machines and Boolean circuits. In fact, neural
nets have an inspiration more grounded in the neurophysiological
structure of the neuronal system.  A survey of the underlying results
concerning the computational power and complexity issues of neuronal
network models can be found in (Sima, \& Orponen, 2003) and references therein.

In a certain sense and from a physical point of view, brain can be
considered as a clock controlled by the internal circadian
rhythm. Hence, it is synchronized with the day/night cycle
(Winfree, 1986).  Roughly speaking, two states can be associated with
this cycle: awake and sleep.  This property is universally observed in
all animals. The cerebral activity is dissociated from the sensory and
motor neurons in the sleep state. This dissociation is not complete
and the brain can still respond to some sensory stimuli.  In fact
there are qualitatively different patterns of neural activity between
different stages of sleep. Basically, two levels, a deepest one and a
shallowest one, alternate during the sleep.  The deepest level of
sleep is attained rapidly and, as sleep progresses, the average level
becomes shallower. The substances that control the connection among
neurons or synopsis monitor theses changes in the neural activity,
which is formed out of composite states occurring in disconnected
brain subdivisions.  When the full connections are restablished, the
waking state of the brain is recovered (Bar-Yam, 1997).

So, as it is suggested by real measurements of the electrical brain activity, 
synchrony seems to be a key concept to explain different aspects of
neuronal behavior.  The activities of two or more neurons, which we
call a {\it functional unit}, are said to be synchronized when some
kind of temporal correlations exists among them.  The conditions for
the emergence of these states are a central issue in the research of
neuronal activity (Borgers, \& Kopell, 2003; Hansel, \& Mato, 2003). 
It has been recently argued
(Eguiluz, Chialvo, Cecchi, Baliki, \& Apkarian, 2003) 
that the distribution of functional connections in the
human brain follows the same distribution of a scale-free network. 
This finding means that
there are regions in the brain that participate in a large number of
tasks while most of the other {\it functional units} are only
involved in a tiny fraction of the brain's activities. The previous
network adds to many examples of such a distribution found in the last
few years in fields as diverse as biological, technological and social
systems (Strogatz, 2001; Dorogovtsev, \& Mendes, 2003; 
Bornholdt, \& Schuster, 2002; Pastor-Satorras, \& Vespignani, 2004). 
They have been termed
scale-free networks because the probability of finding an element with
$k$ connections to other elements of the network follows a
power-law $P(k)\sim k^{-\gamma}$, where $\gamma$ usually lies between
$2$ and $3$. 

In this work we propose a naive approach to mimic the brain
bistability between the sleep and the awake states, and to explain how
to perform the transition between those two basic configurations,
namely the switched on and the switched off states.  In section
\ref{sec1}, a model for a general network showing bistability is
proposed and analyzed. In section \ref{sec2}, the transition between
the active and the non active states is studied. As our model is
thought of as a system made up of functional units and they seem to be
distributed according to a power law, we focus our attention in the
on-off transition for the case of a random scale-free network. The
last section contains our discussion and conclusions.

\section{The Model}
\label{sec1}

Brain is a complex network. Millions of neurons are unidirectionally and
locally interconnected there.  In a first and simple approach one can
consider a {\it functional unit}, i.e. a neuron or group of neurons
(in the following, neuron or functional unit are used indistinctly),
as a discrete dynamical system with two possible states: one active
state and another one with no activity. Let $x_n^i$ be, with
$0<x_n^i<1$, a measurement of the $ith$ network neuron activity at
time $n$.  Take, for instance, a logistic evolution (May, 1976) for the
local neuronal activity:
\begin{equation}
x^i_{n+1} = \bar p\;x^i_n(1-x^i_n).
\label{eq0}
\end{equation}
It presents only one stable state for each $\bar p$. For $\bar p<1$,
the dynamics dissipates to zero, $x_n^i=0$, then it can represent the
functional unit with no activity. For $1<\bar p<4$, the dynamics is non null
and it would represent an active neuron. This local transition is
controlled by the parameter $\bar p$.  The functional dependence of
this local coupling on the neighbor states is essential 
in order to get a good brain-like behavior of the network. It seems reasonable to
take $\bar p$ as a linear function (that we call the {\it Alesves} coupling)
depending on the actual mean value,
$X_n^i$, of the neighboring signal activity and expanding the interval
$(1,4)$ in the form:
\begin{equation}
\bar p = p\;(3X_n^i+1),
\label{eq1}
\end{equation}
with 
\begin{equation}
X_n^i={1\over N_i}\sum_{j=1}^{N_i}x_n^j.
\label{eq2}
\end{equation}
$N_i$ is the number of neighbors of the $ith$ neuron, and $p$,
which gives us an idea of the neuron interaction 
with its first-neighbor neurons, is the control parameter.  
This parameter runs in the range
$0<p<p_{max}$, where $p_{max}\succeq 1$.  Let us observe that there is
an unrealistic bi-directionality in the local neuronal connectivity in
this naive approach to brain-like systems.  This is not a drawback
since networks built under this insight show an interesting
bistability which can mimic the brain behavior. Hence, they present 
an attractive global null configuration that will be identified 
as the {\it turned off} state of the network.  Also they show
a completely synchronized non-null stable configuration that 
we identify as the {\it turned on} state of the network. Thus, 
it is necessary a critical level of noise to
transit from the turned off state to the turned on one for a given
$p$.  The different sleep states, including dreams in human brain,
can be interpreted as a noisy neuronal activity which does not reach
that critical value.  The transition from the awake to the sleep state
can be performed either by decreasing the coupling $p$ or by making
zero the activity of some units. 
All these dynamical properties are universal for different kinds
of local evolution of the same type as equation (\ref{eq0}), the so-called
unimodal maps.\newline
Let us mention at this point that phase synchronization and 
cluster formation in coupled maps on different networks has been studied,
for instance, in (Jalan \& Amritkar, 2003). The results exposed in that work are
very different from those here explained. Concretely, they find that
perfect synchronization leads to clusters with very small number of nodes.
On the contrary, a robust bistability between two 
perfect synchronized states is obtained in our system, 
as it is shown in the next sections.

\subsection{Two-neuron system}

Let us start with the simplest case of two interconnected functional units.
The dynamics is given in this case by the coupled equations:
\begin{eqnarray}
x^1_{n+1} = p\;(3x^2_n+1)\;x^1_n(1-x^1_n), \\
x^2_{n+1} = p\;(3x^1_n+1)\;x^2_n(1-x^2_n).
\label{eq3}
\end{eqnarray}
Depending on the coupling $p$ different dynamical regimes are
obtained (see the details and nomenclature in references
(Lopez-Ruiz, \& Fournier-Prunaret, 2004; 
Lopez-Ruiz, \& Fournier-Prunaret, 2003), 
$p_{max}=1.0843$ in this case):

\begin{itemize}

\item For $0<p<0.75$, the dynamics vanishes.  The two-neuron network
does not have long-term activity. The whole square $[0,1]\times[0,1]$
of initial conditions shrinks to the {\it turned off} configuration,
that is, the fixed point $x_\theta=(0,0)$.

\item For $0.75<p<0.86$, the synchronized state, $x_+=(\bar x, \bar x)$,
with $\bar x={1\over 3}\{1+(4-{3\over p})^{1\over 2}\}$, which arises
from a saddle-node bifurcation for the critical value $p_0=0.75$, is a
stable {\it turned on} state.  This state coexists with $x_\theta$.
The system presents now bistability and depending on the initial
conditions, the final state can be $x_\theta$ or $x_+$. Switching on
the system from $x_\theta$ requires a level of noise in both neurons
sufficient to render the activity on the basin of attraction of $x_+$.
On the contrary, switching off the two-neuron network can be done, for
instance, by making zero the activity of one neuron, or by doing the
coupling $p$ lower than $p_0$.

\item For $0.86<p<0.95$, the active state of the network is now a
period-$2$ oscillation. This new dynamical state bifurcates from $x_+$
for $p=p_c=0.86$.  A smaller noise is necessary to activate the system
from $x_\theta$.  Making zero the activity of one neuron continues to
be a good strategy to turn off the network.

\item For $0.95<p<1$, the active state acquires a new frequency and
presents quasiperiodicity.  It is still possible to switch off the
network by putting to zero one of the neurons.

\item For $1<p<1.03$, bistability is lost. When $p=p_f=1$ the turned
off state $x_\theta$ loses stability and the only stable dynamical state for
$p>p_f$ is now the turned on network.  The network stores the
information in a quasiperiodic state.

\item For $1.03<p<1.08$, a more complex active state is obtained.  In
this range, the network can store more complicated information in the
stable chaotic state, which is now present in the system.

\item For $p>1.0843$, the network loses stability and it can not store
information anymore.
 
\end{itemize}

Let us remark that the two-neuron system exhibits, from a qualitative
point of view, the properties desirable for a brain-like system:
bistability between an active state and another one with no activity
in the range $p_0<p<p_f$ , a necessary noisy level to attain the
activation of the network from the switch off state, and two different
possible strategies to turn off the system from the active state, by
decreasing the synaptic coupling under a critical value or by putting
to zero one of the neurons. \newline We proceed now to show that those
properties are still present in a general complex network.

\subsection{Many neuron system}

The complete synchronization 
(Boccaletti, Kurths, Osipov, Valladares, \& Zhou, 2002) of the network means that $x_n^i=x_n$ 
for all $i$ , with $i=1,2,\ldots,N$ and $N\gg 1$.  Hence, we also have $X_n^i=x_n$. The time
evolution of the network on the synchronization manifold is then given
by the cubic mapping:
\begin{equation}
x_{n+1} = p\;(3x_n+1)\;x_n(1-x_n).
\label{eq4}
\end{equation}  
The fixed points of this system are found by solving $x_{n+1}=x_n$.
The solutions are $x_\theta=0$ and $x_\pm={1\over 3}\{1\pm(4-{3\over
p})^{1\over 2}\}$.  The first state $x_\theta$ is stable for $0<p<1$
and $x_\pm$ take birth after a saddle-node bifurcation for
$p=p_0=0.75$. The node $x_+$ is stable for $0.75<p<1.157$ and the
saddle $x_-$ is unstable.  Therefore bistability between the states
\begin{eqnarray}
x_n^i = x_\theta, & \forall i \longrightarrow & TURNED\;\;\; OFF\;\;\; STATE, \\
x_n^i = x_+, & \forall i \longrightarrow& TURNED\;\;\; ON\;\;\; STATE, 
\label{eq5}
\end{eqnarray}
seems to be also possible for $p>p_0$ in the case of many interacting units.  
But stability in the synchronization manifold does not imply
the global stability.  Small transverse perturbations to this manifold
can make unstable the synchronized states.  Let us suppose then a
general local perturbation $\delta x_n^i$ of the element activity,
\begin{equation} 
x_n^i=x_*+\delta x_n^i,
\label{eq6}
\end{equation}
with $x_*$ representing a synchronized state.  We define the
perturbation of the local mean-field as
\begin{equation} 
\delta X_n^i={3\over N_i}\sum_{j=1}^{N_i}\delta x_n^j.
\label{eq7} 
\end{equation}
If these expressions are introduced in equation (\ref{eq0}), we find
the time evolution of the local perturbations:
\begin{equation}
\delta x_{n+1}^i=p\,(3x_*+1)(1-2x_*)\delta x_n^i+p\,x_*(1-x_*)\delta X_n^i.
\label{eq8}
\end{equation} 
The dynamics for the local mean-field perturbation is derived by
substituting this last expression in relation (\ref{eq7}). We obtain:
\begin{equation}
\delta X_{n+1}^i=p\,(3x_*+1)(1-2x_*)\delta X_n^i+
3p\,x_*(1-x_*){1\over N_i}\sum_{j=1}^{N_i}\delta X_n^j.
\label{eq9}
\end{equation}
We express now the local mean-field perturbations of the first-neighbors 
as function of the local mean-field perturbation $\delta X_n^i$ 
by defining the local operational quantity $\sigma_i^n$,
\begin{equation}
{1\over N_i}\sum_{j=1}^{N_i}\delta X_n^j=\sigma_n^i\;\delta X_n^i,
\label{eq10}
\end{equation}
which is determined by the dynamics itself.  If we
put together the equations (\ref{eq8}-\ref{eq9}), the linear stability
of the synchronized states holds as follows:
\begin{equation}
\left(\begin{array}{c} \delta x_{n+1}^i \\ \delta X_{n+1}^i \end{array} \right) =
\left(\begin{array}{cc} 
p\,(3x_*+1)(1-2x_*) & p\,x_*(1-x_*) \\
0 & p\,(3x_*+1)(1-2x_*)+ 3p\,\sigma_n^i\,x_*(1-x_*)
\end{array} \right)
\left(\begin{array}{c} \delta x_{n}^i \\ \delta X_{n}^i \end{array} \right).
\label{eq11}
\end{equation}
Let us observe that the only dependency on the network topology is
included in the quantity $\sigma_n^i$. The rest of the stability
matrix is the same for all the nodes and therefore it is independent
of the local and global network organization.

The turned off state is $x_*=x_\theta=0$. The eigenvalues of the
stability matrix are in this case $\lambda_1=\lambda_2=p$. Thus, this
state is an attractive state in the interval $0<p<1$.  It loses
stability for $p=1$, then the highest value $p_f$ of the parameter $p$ 
where bistability is still possible satisfies $p_f\leq 1$.

The turned on state $x_+$ verifies $x_*=x_+={1\over 3}\{1+(4-{3\over
p})^{1\over 2}\}$.  If we suppose $\sigma_n^i=\sigma$, the eigenvalues
of the stability matrix are $\lambda_1=2-2p-p(4-{3\over p})^{1\over
2}$ and $\lambda_2=\lambda_1+{\sigma\over 3}(3-2p+p(4-{3\over
p})^{1\over 2})$.  Let us observe that $\lambda_1=-1$ for $p=1$. This
implies that the parameter $p_c$ for which the synchronized state
$x_+$ looses stability verifies $p_c\leq 1$.  Depending on the sign of
$\sigma$, we can distinguish two cases in the behavior of $p_c$:
\begin{itemize}
\item If $0<\sigma<1$, we find that $\mid\lambda_2\mid<1$. Then $x_+$
bifurcates through a global flip bifurcation for $p=p_c=1$. In
this case, the bifurcation of the synchronized state $x_+$ for $p_c=1$
coincides with the loss of the network bistability for $p_f=1$. Hence
$p_c=p_f=1$ for this kind of networks, and the bistability holds
between $x_\theta$ and $x_+$ in the parameter interval
$p_0=0.75<p<p_c=p_f=1$.  As an example, an all-to-all network shows
this behavior because $\sigma=1$. This is represented in the inset of Fig. 1.
\item If $-1<\sigma<0$, then $\lambda_2=-1$ is obtained for a $p=p_c$
smaller than $1$. Therefore it is now possible to obtain an active
state different from $x_+$ in the interval $p_c<p<p_f$.  For instance,
simulations show that the global flip bifurcation of the synchronized
state for a scale free network occurs for $p_c=0.87\pm 0.01$.  A value 
of $p=0.866$ is
obtained from the stability matrix by taking $\sigma=-1$.  For this
particular network it is also found that $p_f=1$. Then, bistability is
possible in the range $p_0=0.75<p<p_f=1$ for this kind of
configuration.  But now an active state with different dynamical
regimes is observed in the interval $p_c=0.87<p<p_f=1$. If we identify
the capacity of information storing with the possibility of the system
to access to complex dynamical states, then, we could assert, in this
sense, that a scale free network has the possibility of storing more
elaborated information in the bistable region that an all-to-all
network.
\end{itemize}
Let us note that $\sigma$ also indicates a different behavior of local
dissipation, as expression (\ref{eq10}) suggests.  A positive $\sigma$
means a local in-phase oscillation of the node signal and mean-field
perturbations. A negative $\sigma$ is meaning a local out of phase
oscillation between those signal perturbations. Hence, $\sigma$ also brings
some kind of structural network information.  In all the cases the
stability loss of the completely synchronized state is mediated by a
global flip bifurcation. The new dynamical state arising from that
active state for $p=p_c$ is a periodic pattern with a local period-$2$
oscillation. The increasing of the coupling parameter monitors other
global bifurcations that can lead the system towards a pattern of local
chaotic oscillations.

\section{Transition between On-Off States}
\label{sec2}

We proceed now to show the different
strategies for switching on and off a random scale free network. 
The choice of this network is suggested by the recent work
(Eguiluz et al., 2003; Buzs\`aki, Geisler, Henze, \& Wang, 2004) 
on the connections distribution among functional units in brain. 
They find it to be a power-law distribution.
Following this insight, we generate a scale-free network following the
Barab\'asi-Albert (BA) recipe (Barabasi, \& Albert, 1999). 
In this model, starting
from a set of $m_0$ nodes, one preferentially attaches each time step
a newly introduced node to $m$ older nodes. The procedure is repeated
$N-m_0$ times and a network of size $N$ with a power law degree
distribution $P(k)\sim k^{-\gamma}$ with $\gamma=3$ and average
connectivity $\langle k \rangle=2m$ builds up. This network is a clear
example of a highly heterogenous network, in that the degree distribution
has unbounded fluctuations when $N\rightarrow\infty$. The exponent reported for the brain
functional network has $\gamma < 3$. However, studies of percolation
and epidemic spreading (Pastor-Satorras, \& Vespignani, 2004; 
Callaway, Newman, Strogatz, \& Watts, 2000; 
Moreno, Pastor-Satorras, \& Vespignani, 2002; Vazquez, \& Moreno, 2003) 
on top of
scale-free networks has shown that the results obtained for $\gamma=3$
are consistent with those corresponding to lower values of $\gamma >
2$. Therefore, we expect that the results shown henceforth are not
biased by the use of a different exponent. As
explained before, network bistability between the active and non active
states is here possible in the interval $p_0=0.75<p<p_f=1$ (Fig. 1).

\subsection{Switching off the network}

Two different strategies can be followed to carry the network from the
active state to that with no activity (Fig. 1).
\begin{itemize}
\item {\it Route I}: By doing the coupling $p$ lower than
$p_0$.  This is the easiest and more natural way of performing such an
operation.  In our brain-like interpretation, it could represent the
decrease (or increase, it depends on the specific function)
of the synaptic substances  that provokes the transition
from the awake to the sleep state. The flux of these chemical
activators is controlled by the internal circadian clock, which is
present in all animals, and which seems to be the result of living
during millions of years under the day/night cycle.
\item {\it Route II}: By switching off a critical fraction of neurons
for a fixed $p$.  This is done by looking over all the elements of the
network, and considering that the element
activity is set to zero with probability $\lambda$ 
(which implies that on average $\lambda N$
elements are reset to zero). The result of this operation is shown in
Fig. 2. Here, it is plotted for different $p$'s the relative size of
the biggest (giant) cluster of connected active nodes in the network
versus $\lambda$. Note that this procedure does not take into account
the existence of connectivity classes, but all nodes are equally
treated. The procedure is thus equivalent to simulations of random
failure in percolation studies (Callaway et al., 2000). The strategy in which highly connected
functional units are first put to zero is more aggressive and leads to
quite different results. 

Each curve presents three different zones depending on $\lambda$:
\begin{quote}
\item - the {\it robust phase}: For small $\lambda$, the network is
stable and only those states put to zero have no activity.  There is a
linear dependence on the giant cluster size with $\lambda$.  In this
stage, the switched off nodes do not have the capacity to transmit its
actual state to its active neighbors.
\item - the {\it weak phase}: For an intermediate $\lambda$, the nodes
with null activity can influence its neighborhood and switch off some
of them. The linearity between the size of the giant cluster and
$\lambda$ shows a higher absolute value of the slope than in the
robust zone.
\item - the {\it catastrophic phase}: When a critical $\lambda_c$ is
reached, the system undergoes a crisis. The sudden drop in this zone
means that a small increase of the non active nodes leads the system to
a catastrophe; that is, the null activity is propagated through all
the network and it becomes completely down.
\end{quote}
\end{itemize}
It is worth noticing that when the system is outside the bistability
region for $p>1$, the catastrophic phase does not take place. Instead,
the turned off nodes do not spread its dynamical state and the
neighboring nodes do not die out. This is because the dynamics of
an isolated node is self-sustained when $p>1$. Consequently, we observe that the
network breaks down in many small clusters and the transition
resembles that of percolation in scale free nets 
(Callaway et al., 2000; Vazquez, \& Moreno, 2003).

\subsection{Switching on the network}

Two equivalent strategies can be followed for the case
of turning on the network (Fig. 3): (I) For a fixed $p$, we can increase
the maximum value $\epsilon$ of the noisy signal, which is randomly
distributed in the interval $(0,\epsilon)$ over the whole system.
When $\epsilon$ attains a critical value $\epsilon_c$, the noisy
configuration can leave the basin of attraction of $x_{\theta}$,
which seems to have the form in phase space 
of a ``hollow cane'' ({\it canuto}) around it, and then 
the network rapidly evolves toward the turned on state; 
(II) If this
operation is executed by letting $\epsilon$ to be fixed and by
increasing the coupling parameter $p$, the final result of switching on
the network is reached when $p$ takes the value for which
$\epsilon=\epsilon_c$. The final result is identical in both cases.

Let us remark that the strategy equivalent to the former Route II is
not possible in this case. It is a consequence of the fact that a
switched off neuron can not be excited by its neighbors and it will
maintain indefinitely the same dynamical state ($x_i=0$).

\section{Conclusions}
 
One of the most challenging scientific problems today is to understand
how the millions of neurons of our brain give rise to the emergent
property of thinking. Different aspects of neurocomputation take
contact on this problem: how brain stores information and how brain
processes it to take decisions or to create new information. These are
characteristics more or less accepted and observed in all the brains.

Other universal properties of this system are more evident.  
One of them is the existence of a regular daily behavior:
the awake and the sleep.  The internal circadian rhythm is
closely synchronized with the cycle of sun light.  Roughly speaking and
depending on the particular species, the brain is awake during the day
and it is slept during the night, or vice versa.  Hence, this
evident bistability does not depend on the precise architecture of a
special brain.

In this work, we have studied a general network with local logistic
dynamics that presents global bistability between an active
synchronized state and another synchronized state with no activity.
This property is topology and size independent.  This is a
direct consequence of the local mean-field multiplicative coupling
among the first-neighbors (the {\it Alesves} coupling). 
Different routes to transit from one state
to the other have been explored for the important case of a
scale free network.  If a formal relationship is established between
the switched on and switched off states of that network, and
the awake and sleep states of a brain, respectively, one would be
tempted to assert that this model is a good qualitative representation
for explaining that specific bistable
behavior. Other analogies could be suggested in reference to the
usual functioning and the failures of a power line, or also
the brain bistability in pattern recognition is another intriguing
neural phenomenon. Furthermore, we are convinced that this
model, regardless of its simplicity, can bring new qualitative insights 
on how the brain works.

%\section{Conclusions}

\newpage
\begin{center} {\bf Figures} \end{center}

\begin{figure}[h]
\begin{center}
\epsfig{file=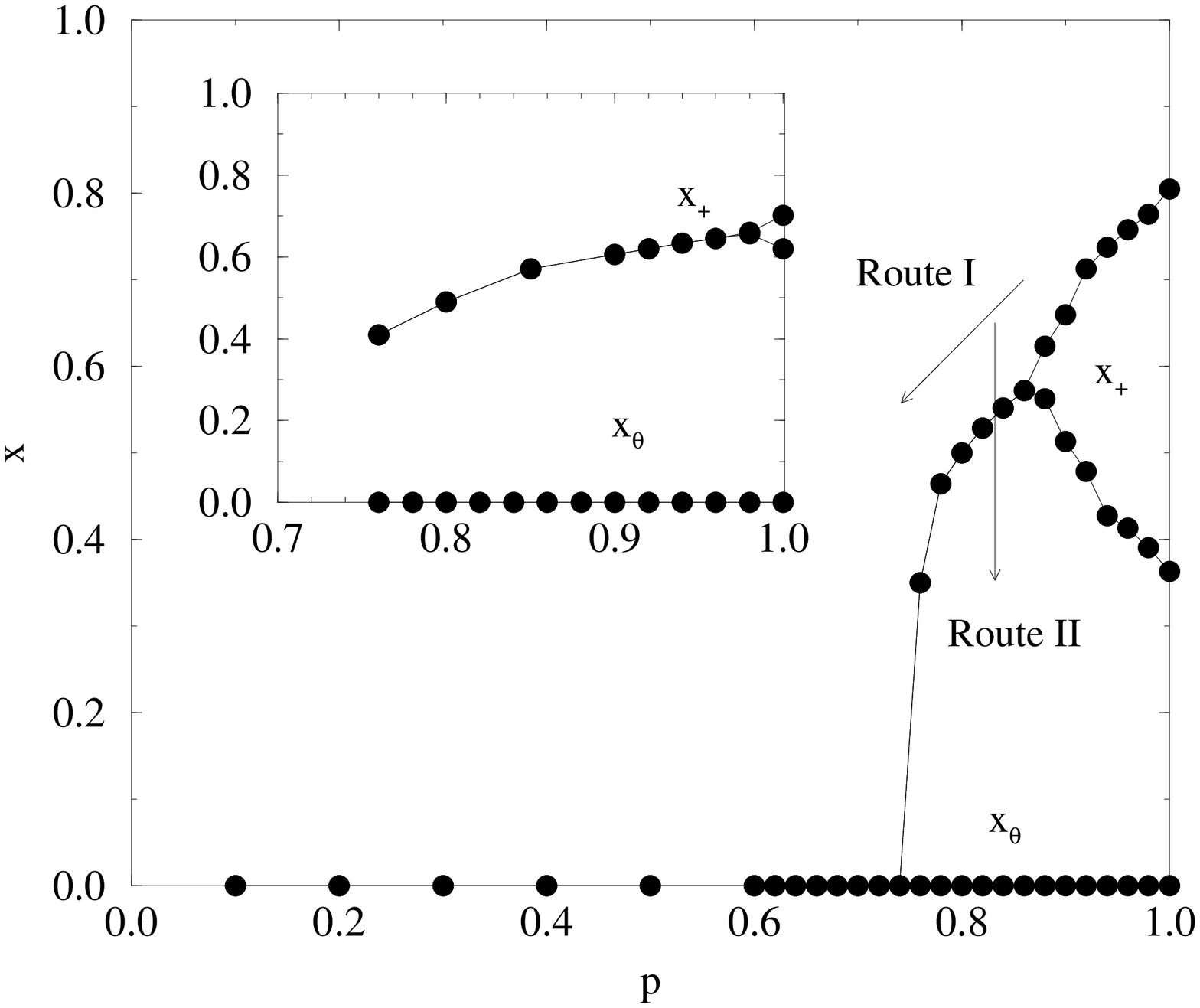,width=3.6in,angle=-90,clip=1}
\end{center}
\caption{Stable states $(x_{\theta},x_+)$ of the network for
$0<p<1$.  Let us observe the two zones of bistability: $p_0<p<p_c$ and
$p_c<p<p_f$.  The main figure correspond to a scale free network made
up of $N=10^4$ elements: $p_0=0.75$, $p_c=0.87\pm 0.01$ and $p_f=1$. The
inset shows the same graph but in an all-to-all network of the same
size: $p_0=0.75$, $p_c=p_f=1$. Initial conditions for the $x_i$'s were
drawn from a uniform probability distribution in the interval $(0,1)$.}
\label{fig1}
\end{figure}

\begin{figure}[t]
\begin{center}
\epsfig{file=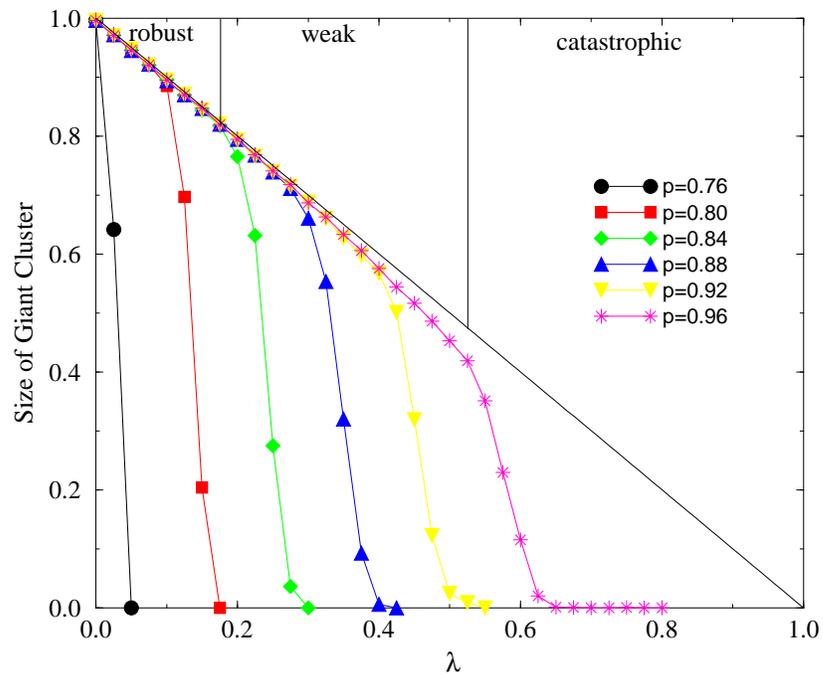,width=3.6in,angle=-90,clip=1}
\end{center}
\caption{Turning off a scale free network. Three different phases
in the behavior of the giant cluster size versus $\lambda$ (fraction
of switched off nodes) are observed. These three phases are illustrated 
for $p=0.96$: the robust phase, the weak phase and the catastrophic 
phase (see the text). Other network parameters are as those of Fig.\ \ref{fig1}.}
\label{fig2}
\end{figure}

\begin{figure}[t]
\begin{center}
\epsfig{file=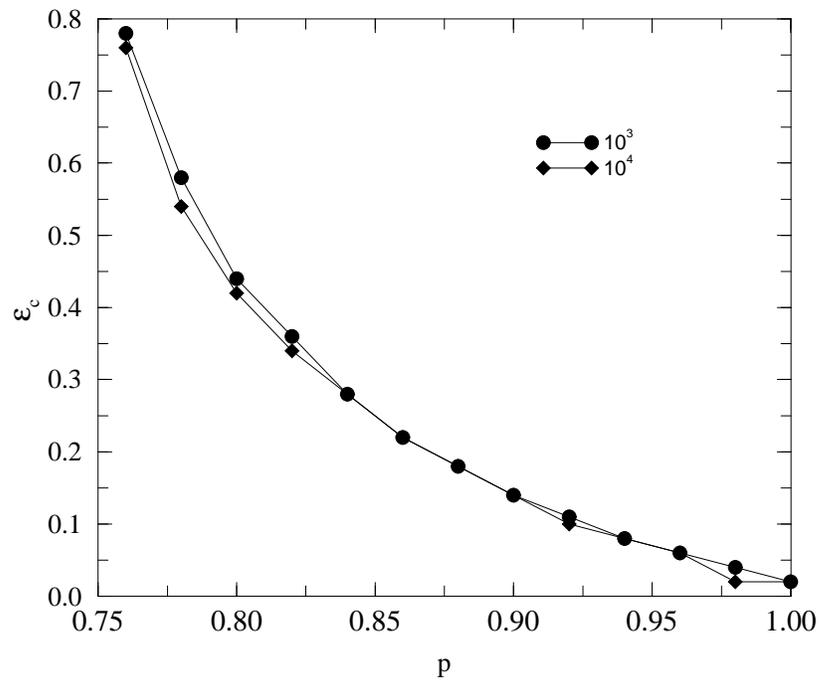,width=3.6in,angle=-90,clip=1}
\end{center}
\caption{Turning on a scale free network. For a fixed $p$, a noisy
signal randomly distributed in the interval $(0,\epsilon)$ is
assigned to every node. When $\epsilon$ reaches the
critical level $\epsilon_c$ the network becomes switched
on. Other network parameters are as those of Fig.\ \ref{fig1}.}
\label{fig3}
\end{figure}

\end{document}